\documentclass{article}
\usepackage{spconf,amsmath,graphicx}
\usepackage{booktabs}

\title{Towards Unsupervised Single-Channel Blind Source Separation using Adversarial Pair Unmix-and-Remix}
\name{Yedid Hoshen}
\address{Hebrew University of Jerusalem and Facebook AI Research}
\begin{document}
%
\maketitle
\begin{abstract}
Blind single-channel source separation is a long standing signal processing challenge. Many methods were proposed to solve this task utilizing multiple signal priors such as low rank, sparsity, temporal continuity etc. The recent advance of generative adversarial models presented new opportunities in signal regression tasks. The power of adversarial training however has not yet been realized for blind source separation tasks. In this work, we propose a novel method for blind source separation (BSS) using adversarial methods. We rely on the independence of sources for creating adversarial constraints on pairs of approximately separated sources, which ensure good separation. Experiments are carried out on image sources validating the good performance of our approach, and presenting our method as a promising approach for solving BSS for general signals. 
\end{abstract}
\section{Introduction}
\label{sec:intro}

The task of single-channel blind source separation (BSS) sets to reconstruct each of several sources (typically additively) mixed together. The task is poorly determined as more information needs to be reconstructed than the number of observations. BSS methods therefore need to rely on strong signal priors in order to constrain source reconstruction. Many priors were proposed for this task each giving rise to different optimization criteria. Source priors include: sparsity in time-frequency, non-Gaussian distribution of sources and low rank of sources. Recently, deep neural network methods that learn high quality signal representations (a form of prior learning) made much progress on single-channel source separation for cases where clean samples of each of the sources were available in training. This allowed creating synthetically mixed datasets, where random clean samples from each source are sampled and additively mixed. A deep neural network is then used to regress each of the components from the synthetic mixture. Such approaches are very effective due to learning source priors, rather than using generic hand-specified priors. Recent work was carried out to reduce the supervision required to having clean samples of only a single source, however when only mixed source samples are available (and no clean samples), classical methods are still used.

In this paper, we introduce a machine learning-based approach for the single-channel BSS case i.e. when no clean source samples are available at training time. Our method is based on generative adversarial networks (GANs) and uses a mixture of distributional, energy and cycle constraints to achieve high-quality unsupervised source separation. Our method makes the assumption of distributional independence between sources. In this work, we concentrate on the case where we are given mixed images (which is similar to having short audio clips) and do not take into account temporal priors (e.g. HMM models), which are left to future work. Our method is experimentally shown to outperform state-of-the-art single-channel BSS methods for image signals. Due to the strong performance on image signal separation, we believe that our approach presents a novel and promising direction for solving the long-standing task of single-channel BSS for general signals.

\section{Previous Work}
\label{sec:prev}

Single-channel BSS has received much attention. The best results are typically obtained by using strong priors about the signals. Robust-PCA \cite{huang2012singing} separates instrumental and vocal sources by assuming that one source is low-rank while the vocal source is sparse. Results are improved with supervision \cite{chan2015vocal}. For repetitive signals, Kernel Additive Models \cite{liutkus2014kernel} may be used. Using temporal continuity was exploited by Roweis \cite{roweis2001one} and Virtanen \cite{virtanen2007monaural}. Work was also done on designing priors for image mixture separation e.g. Levin and Weiss \cite{levin2004separating} used the presence of corners, although this method required access to clean signals.

\begin{figure*}[t]
  \centering

\includegraphics[width=0.45\linewidth]{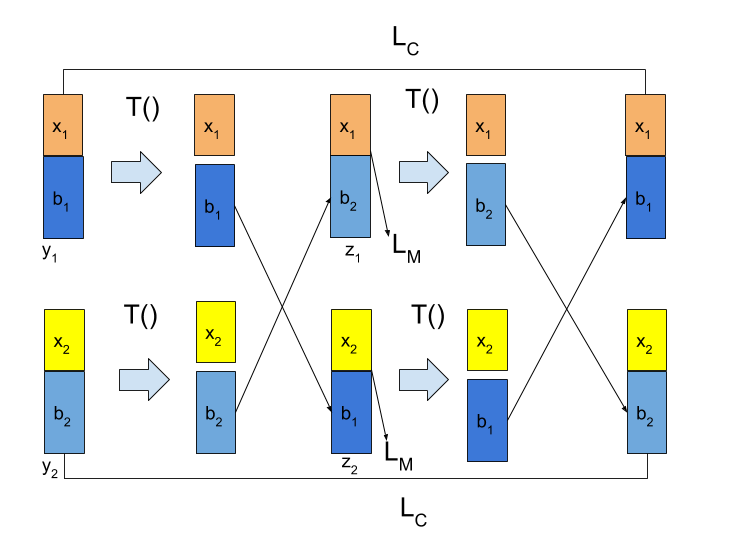}

  \caption{A schematic of our architecture: we select a pair of samples $y_1$ and $y_2$, and separate them into estimated sources. We then flip the combination of sources between the two signals to synthesize $z_1,z_2$. We optimize $T(y)$ (implemented as $y \cdot M(y))$ to make the new mixtures $z_1,z_2$ indistinguishable from the original mixtures $y_1,y_2$. We then separate and remix the new mixtures again. The optimal separation function will recover the original mixtures $y_1,y_2$.  }
  \label{fig:gen_shoes}
\end{figure*}

Machine learning methods take away some of the difficulty in manual prior design. Speaker source separation was achieved by deep neural networks by \cite{huang2014deep}. Permutation-invariant training \cite{yu2017permutation} allows speaker independent separation using deep learning methods. The above methods were used in a supervised source separation context i.e. when clean samples of each source are available. Supervised deep methods were also used for image separation (e.g. in the context of reflection removal \cite{chi2018single,zhang2018single}). Generative adversarial networks (GANs) \cite{goodfellow2014generative} were used by some researchers in a supervised setting, for learning a better loss function \cite{stoller2017adversarial, subakan2017generative}, typically with modest gains. They have been similarly used in image separation tasks with more significant gains \cite{chi2018single}. 

There have been few attempts to apply deep methods for the unsupervised regime. In a recent work \cite{anon2018neural}, we proposed using deep learning methods for semi-supervised separation (when samples of one source are available but not of the other). One of our baselines proposed a GAN-based method for learning the masking function. This method was outperformed by our main method - Neural Egg Separation which is non-adversarial. In this paper, we deal with the more challenging scenario, where no clean examples are available for any of the sources.

The architecture in DRIT \cite{lee2018diverse} used for image style and content disentanglement bears some relation to ours as it uses cycle and pair-adversarial constraints . Our approach is different in key ways: we operate in the input rather than latent domain, we use masking rather than a set of encoders significantly constraining the network and improving results. We also introduce the energy equity term which is critical for the success of our approach.

\section{Adversarial Unmix-and-Remix}
\label{sec:method}

In the following, we denote the set of mixed signals as $y_1,y_2..y_N$. For ease of explanation, our formulation will assume two sources (however in Sec.~\ref{sec:discuss} we explain why there is no loss of generality). We name our sources, $\cal X$ and $\cal B$ such that every mixed signal $y_i$ consists of separate sources $x_i$ and $b_i$, but no examples of such sources are given in the training set. 

Our objective is to learn separation function $T()$ which separates a mixed signal $y$ into its sources $x$ and $b$. We parametrize the separation function by a multiplicative masking operation $M()$ as shown in Eq.~\ref{eq:mask}:

\begin{equation}
\label{eq:mask}
T(y) = y \cdot M(y)
\end{equation}

The separated sources are therefore given by $y \cdot M(y)$ and $y \cdot (1-M(y))$. The masking function is learned as part of training.

Our method begins by sampling two mixed signals, which we will denote $y_1$ and $y_2$ (with no loss of generality). We operate the masking function on each mixture obtaining:

\begin{equation}
\label{eq:pair_sep}
\begin{array}{c}
\widetilde{x_1} = T(y_1)~~~~~ \widetilde{b_1} = y_1 - T(y_1) \\
\widetilde{x_2} = T(y_2)~~~~~ \widetilde{b_2} = y_2 - T(y_2)
\end{array}
\end{equation}

We make the assumption of independence between the two sources $\cal X$ and $\cal B$. This assumption is valid for many interesting mixtures of signals such as images and reflections or foreground and background noise.

With this assumption, we can now synthesize new mixed signals $z_1$ and $z_2$, which are obtained by flipping the source combinations between the two pairs:

\begin{equation}
\label{eq:synth}
\begin{array}{c}
z_1 = \widetilde{x_1} + \widetilde{b_2} \\
z_2 = \widetilde{x_2} + \widetilde{b_1}
\end{array}
\end{equation}
 
Although the new mixed signals will be different from $y_1$ and $y_2$, we make the observation that their \textit{distribution} should be the same as that of $y_1$ and $y_2$, if the separation works correctly. Therefore to encourage correct separation, we require the distribution of $\cal Y$ and $\cal Z$ to be identical. This can be enforced using an adversarial domain confusion constraint. Specifically this works by training a discriminator $D()$ to attempt to identify if a specific signal comes from $\cal Y$ or from $\cal Z$. The discriminator is trained using the following LS-GAN \cite{mao2017least} loss function:

\begin{equation}
\label{eq:disc}
arg\min_D{L_D} = \sum_{y \in \cal Y}(D(y) - 1) ^ 2 + \sum_{z \in \cal Z}D(z)^2
\end{equation}

We co-currently train the masking function $M()$ so that it acts to fool the discriminator by making the mixed signals $\cal Z$ as similar as possible to $\cal Y$:

\begin{equation}
\label{eq:confusion}
arg\min_M{L_M} = \sum_{z \in \cal Z}(D(z) - 1) ^ 2
\end{equation}

Where $z$ iterates over all $z_1$ and $z_2$. 

Although perfect separation is one possible solution of the distribution matching equation, another acceptable by unwanted solution is $\tilde{x} = y$ and $\tilde{b} = 0$. This trivial solution satisfies the distributional matching perfectly, but obviously achieves no separation. To combat this trivial solution, we add another loss term which favors solutions that give non-zero weights to the different sources:

\begin{equation}
\label{eq:energy}
L_E = \sum_{y \in \cal Y} (y \cdot M(y))^2 + (y \cdot (1-M(y)))^2
\end{equation}

A further constraint on the separation can be obtained by another application of the separation function of the synthetic mixture signal pair $z_!$ and $z_2$. We perform the same unmixing and remixing operation as performed in the first stage: 
\begin{equation}
\label{eq:pair_synth_sep}
\begin{array}{c}
\overline{x_1} = T(z_1)~~~~~ \overline{b_2} = z_1 - T(z_1) \\
\overline{x_2} = T(z_2)~~~~~ \overline{b_1} = z_2 - T(z_2)
\end{array}
\end{equation}

In this case, we notice that the result should be identical to the original unmixed signals $y_1$ and $y_2$:

\begin{equation}
\label{eq:cycle_synth}
\begin{array}{c}
\overline{y_1} = \overline{x_1} + \overline{b_1} \\
\overline{y_2} = \overline{x_2} + \overline{b_2}
\end{array}
\end{equation}

We therefore introduce a "cycle" loss term, ensuring that the double application of unmixing and remix operation of a pair of mixed signals recovers the original signals:

\begin{equation}
\label{eq:cycle_loss}
L_C = \sum_{y \in \cal Y} \|\overline{y}, y\|
\end{equation}

To summarize, our method optimizes the separation function $T()$ (which is implemented using multiplicative masking function $M()$ as described in Eq.~\ref{eq:mask}). The loss function to be optimized is the combination of the domain confusion loss $L_M$, the energy equity loss $L_E$ and the cycle reconstruction loss $L_C$:

\begin{equation}
\label{eq:cycle_loss}
arg\min_M{L_{Total}} = L_C + \alpha \cdot L_M + \beta \cdot L_E
\end{equation}

We also adversarially optimize the discriminator $D()$ as described in Eq.~\ref{eq:disc}. 

\begin{figure*}

  \centering
  
	\caption{A Qualitative Comparison of MNIST and Shoes/Bags Separation \label{fig:qual}}
    \begin{tabular}{cccccccccc}
    \toprule
    Mix & RPCA & GLO & Ours & GT & Mix & RPCA & GLO & Ours & GT \\
    \midrule
    \includegraphics[width=0.076\linewidth]{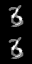} &
    \includegraphics[width=0.076\linewidth]{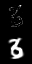} &
    \includegraphics[width=0.076\linewidth]{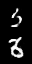} &
    \includegraphics[width=0.076\linewidth]{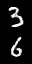} &
    \includegraphics[width=0.076\linewidth]{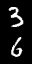} &
    \includegraphics[width=0.076\linewidth]{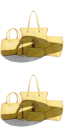} &
    \includegraphics[width=0.076\linewidth]{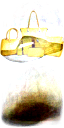} &
    \includegraphics[width=0.076\linewidth]{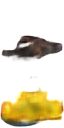} &
    \includegraphics[width=0.076\linewidth]{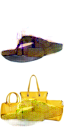} &
    \includegraphics[width=0.076\linewidth]{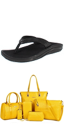}
    \end{tabular}
\end{figure*}

\section{Implementation}
\label{sec:imp}

We implemented the masking function $M()$ by an architecture that follows DiscoGAN \cite{kim2017learning} with 64 channels (at the layer before last, each preceding layer having twice the number of channels). The discriminator followed a standard DCGAN \cite{dcgan} architecture with 64 channels. We used a learning rate of $0.0001$. Optimization was carried out by SGD with the ADAM update rule. We carried out $4$ mask update steps for every $D()$ update. We used $\alpha=5$ for the adversarial loss $L_M$ and $\beta=5$ for the energy equity loss $L_E$.

\section{Evaluation}
\label{sec:eval}

In this section, we evaluate the effectiveness of our method for image separation tasks against other state-of-the-art unsupervised single channel source separation methods.

\textbf{Datasets:} We use the following image datasets in our experiments:

\textit{MNIST:} The MNIST dataset \cite{mnist} consists of $50000$ training and $10000$ validation images of hand written digits $0-9$. The images are roughly evenly distributed between the different classes. The original image resolution is $28 \times 28$. In order to use standard generative architectures, we pad the images by $2$ pixels from each direction to have a size of $32 \times 32$. We split the dataset into two sources: the images of the digits from $0-4$ and the images of the digits from $5-9$. A random image is sampled from each source, and then combined with equal weights. We sampled $25k$ training mixture images (from the training sets), and $5k$ validation images from the validation set.

\textit{Shoes and Bags:} The Shoes dataset \cite{edges2shoes} first collected by Yu and Grauman consists of color images of different types of shoes. We rescale the image resolution to $64 \times 64$. The Handbags dataset \cite{edges2bags} collected by Zhu et al. consists of color images of a variety of handbags. We also rescale these images to a resolution of $64 \times 64$. The two datasets are often used in image generative modeling tasks. As masking works better when the background has $0$ value, we run our experiments on the inverted intensity images (i.e. from image $I$, we use $255 - I$). Our sampling procedure is to randomly sample a shoe image and a handbag image (without replacement) and mix them with equal weights. This is repeated $10k$ times to form our training set. We similarly sample $5k$ mixture test images. No source image is repeated between the train and test sets. 

\textbf{Methods:} Separating two images from arbitrary image classes does not satisfy the requirements for any of the typical priors as there are no obvious temporal, sparsity or low-rank constraints. We compare against RPCA \cite{huang2012singing} which is representative of methods that use strong priors. To represent de-compositional methods we compare against GLO, a generative model (which in \cite{anon2018neural} was preferable to NMF). To have an upper bound for the quantitative comparison, we also give the fully supervised performance (using the same masking function architecture that we used). We stress however that our method is fully unsupervised, and we do not expect to do better than the fully supervised method. 

\textbf{Qualitative Results:} A qualitative comparison is presented in Fig.~\ref{fig:qual}. We observe that RPCA completely fails on this task, as the sparse/low-rank prior is not suitable for arbitrary images. GLO tended to result in uneven separation - one generator containing a part of one source, while the other generator containing a mixture of the sources. Our method, generally resulted in clean separation of the sources. In highly textured regions, we sometimes saw some "dripping" of the texture to the other source.

\textbf{Quantitative Results:} We present a quantitative comparison on MNIST and Shoes/Bags . The metrics are PSNR (in Tab.~\ref{tab:results_psnr}) and SSIM \cite{wang2003multiscale} (in Tab.\ref{tab:results_ssim}), which are standard image reconstruction quality metrics. In both cases we can observe that GLO performed much better than RPCA (due to the prior in RPCA being unsuitable for this more general task). Our method far outperformed both baseline methods, due to our careful separation design. The performance of our method approaches the supervised separation performance, however there still is a significant performance gap due to supervision, which is unsurprising. In ablation experiments, we found that the adversarial loss and the energy equity loss were essential for the convergence of our method to the correct solution. The cycle constraint was found to only slightly increase stability of convergence and did not increase accuracy. Overall, we can conclude that the results validate the strong performance of our method for separating image sources.  

\begin{table}[t]
  \centering
      
  \caption{Separation Accuracy (PSNR)  \label{tab:results_psnr}}

    \begin{tabular}{lcccc}
    \toprule
   $Dataset$ & \textit{RPCA} & \textit{GLO} & \textit{Ours} & \textit{Sup}    \\   
    \midrule
     MNIST & 11.5 & 13.0 & \textbf{20.4} & 24.4  \\
     Shoes and Bags & 7.9 & 12.0 & \textbf{19.0} & 22.9 \\
	 \bottomrule
    \end{tabular}
\end{table}

\begin{table}[t]
  \centering
      
  \caption{Separation Accuracy (SSIM) \label{tab:results_ssim}}

    \begin{tabular}{lcccc}
    \toprule
   $Dataset$ & \textit{RPCA} & \textit{GLO} & \textit{Ours} & \textit{Sup}    \\   
    \midrule
     MNIST & 0.36 & 0.74 & \textbf{0.90} & 0.96  \\
     Shoes and Bags & 0.18 & 0.51 & \textbf{0.73} & 0.86 \\
	 \bottomrule
    \end{tabular}
\end{table}

\section{Discussion}
\label{sec:discuss}

We make several comments about our work:

\textbf{Priors:} Our method was shown to be effective at separating mixtures of images, using no image specific priors such as repetition, sparsity or low-rank. It is therefore potentially extensible to all 2D signals. We make the general assumption that the distributions of the two signals are independent.

\textbf{Spectrograms:} Preliminary experiments on spectrograms were not able to match the success of the method for image separation. We think that this is due to GAN modeling of images being more developed than that of spectrograms. We believe that with future progress in adversarial architecture for spectrograms, our technique will be able to separate audio clips.

\textbf{Multiple Sources:} Although the formulation in this work only dealt with $2$ sources, it can be applied to a larger number of sources, by a applying the method in a binary tree-like structure (recursively applying our method on each separated "source" until reaching the leaves - the clean sources). We note however that the binary tree-like structure will need to have a stopping criterion detecting when a clean source has been found (similar to a leaf in a tree). We leave this to future work.

\section{Conclusion}
\label{sec:conc}

In this paper, we introduced a novel method for the single-channel separation of sources without seeing any clean examples of the individual sources. Previous methods have been able to achieve this either by learning strong priors from clean data or by carefully hand-crafting priors for particular sources. Our method makes very few assumptions on the sources, making it applicable to signals for which strong priors are not known.  We demonstrated that our method works well on separating mixtures of images. Future work on adversarial training for spectrograms is needed to extend our approach to audio sources.

\bibliographystyle{IEEEbib}
\bibliography{Template}

\end{document}